# Maxwell’s equations for a mechano-driven, shape-deformable, charged-media system, slowly moving at an arbitrary velocity field $v(\boldsymbol{r},t)$

PAPER

# Maxwell's equations for a mechano-driven, shape-deformable, charged-media system, slowly moving at an arbitrary velocity field *v*(*r*,t)

**Zhong Lin Wang**[1,2,3]

[1] Beijing Institute of Nanoenergy and Nanosystems, Chinese Academy of Sciences, Beijing 101400, People's Republic of China
[2] School of Materials Science and Engineering, Georgia Institute of Technology, Atlanta, Georgia 30332-0245, United States of America
[3] College of Nanoscience and Technology, University of Chinese Academy of Sciences, Beijing 100049, People's Republic of China

**E-mail:** zhong.wang@mse.gatech.edu

**Keywords:** Maxwell's equations for mechano-driven slow-moving media, Galilean electromagnetism, triboelectric nanogenerators

## Abstract

The differential form of the Maxwell's equations was first derived based on an assumption that the media are stationary, which is the foundation for describing the electro-magnetic coupling behavior of a system. For a general case in which the medium has a time-dependent volume, shape and boundary and may move at an arbitrary velocity field $\boldsymbol{v}(\boldsymbol{r},\ t)$ and along a general trajectory, we derived the *Maxwell's equations for a mechano-driven slow-moving media system* directly starting from the integral forms of four physics laws, which should be accurate enough for describing the coupling among mechano-electro-magnetic interactions of a general system in practice although it may not be Lorentz covarance. Our key point is directly from the four physics laws by describing all of the fields, the space and the time in the frame where the observation is done. The equations should be applicable to not only moving charged solid and soft media that has acceleration, but also charged fluid/liquid media, e.g., fluid electrodynamics. This is a step toward the electrodynamics in non-inertia frame of references. General strategies for solving the Maxwell's equations for mechano-driven slowing moving medium are presented using the perturbation theory both in time and frequency spaces. Finally, approaches for the electrodynamics of moving media are compared, and related discussions are given about a few interesting questions.

## 1. Introduction

Maxwell's equations are probably the most important equations for the field of physics, which have huge importance in both fundamental science and practical technologies [1]. Starting from experimentally observed physics laws, such as Lenz's law and Ampere's law, the exact mathematical expressions of them are given in differential form with proper boundary conditions. Although the most commonly used Maxwell's equations are for media with fixed volumes, boundaries and at stationary, such an assumption is rarely mentioned in textbook, such as Jackson's book on '*Classical Electrodynamics*' [2], so that generations of students and scientists may not realize the preconditions under which the Maxwell's equations in differential forms were derived. Therefore, we always believe in mind that Maxwell's equations should be applicable to any case in electrodynamics.

Historically, Maxwell first added into the induction equation the term that described the induction due to the motion of the medium in 1865 [3]. Later, Hertz systematically extended Maxwell's theory for moving media in 1890 [4], but his equations were valid only for conductors and needed to be expanded on the cases of dielectrics and empty space. Minkowski derived electrodynamic equations for moving media using the principle of relativity in 1908 [5]. Then, the development of electrodynamics for moving media was rather slow due to the appearance of theory of relativity.

In the last 50 years, the interest on the study of electrodynamics of moving media has been revived from time to time [6, 7]. There are a number of studies about the Maxwell's equations for slow-moving media/bodies ($v \ll c$) with a focus on the scattering, reflection and transmission of electromagnetic waves from moving media [8–11]. The pioneer work of Le Belllac and Levy-Leblond started the work on Galilean Electromagnetism [12]. A comprehensive review on the topic has been given by Rousseaux on the electromagnetism theory developed under Galilean transformation [13], who concluded: 'Galilean Electromagnetism is not an alternative to Special Relativity but is precisely its low-velocity limit in Classical Electromagnetism', establishing the validity of the approach toward engineering and physics applications. This work also sets the foundation for our following work.

Most of the existing work assumes that the moving velocity of the media is a constant and along a straight line, e.g., inertia frame, especially for special relativity. For a medium that moves along a complex trajectory, there is no simple approach of using Lorentz transformation! Starting from the integral forms of the four physics laws, we have systematically studied the Maxwell's equations for a media that has a fixed boundary and volume and is moving at a slow-speed as a *solid translation* [14], which means that the moving velocity $\boldsymbol{v}(t)$, including both direction and amplitude, is time-dependent but space-independent, e.g., a solid translation along an arbitrary trajectory. General theories, associated analytical solutions as well as potential applications have been developed for the expanded Maxwell's equations. Although our derivation was based on the conditions assumed for Galilean transformation (note: Galilean transformation is for a medium that moves at a constant velocity along a straight line trajectory), the validity of the approach has been examined under various conditions. It is this work that has inspired new interest in the field. An alternative equivalent (traveling wave-like) description of Maxwell's equations is derived in [15], which can reduce to the (extended) Hertz-form equations from non-relativistic (low speed) expansion, which are for the fields in the Lab frame but use the coordination in the co-moving frame $\boldsymbol{r}'$. Possible superlumimal behavior in the media will not cause inconsistency because only the leading terms in the non-relativistic expansion are included in the Hertz (and extended Hertz) equations. The superluminal behavior will not emerge if all terms in the non-relativistic expansion are taken into account. Li *et al* [16] have proven that, from the viewpoint of different reference frames, the results of [14] are valid, especially when there are many moving media with different speeds, redefining the physical quantities of different co-moving frame to Lab one will simplify the solving process greatly. So, Li *et al* concluded that our work of [14] is really valuable in practical applications such as triboelectric nanogenerators. Sheng *et al* [17] have made a systematic description on the low-speed limit of Lorentz transformed Maxwell's equations starting from the field theory, and they have pointed out that the following condition is also required in order to reach the results derived by Wang [14]: $|\boldsymbol{v} \cdot \frac{\partial}{\partial t}\boldsymbol{F}|$ for $\boldsymbol{F} = \boldsymbol{B}$ or $\boldsymbol{D}$, or $|\boldsymbol{v} \times \frac{\partial}{\partial t}\boldsymbol{F}|$ for $\boldsymbol{F} = \boldsymbol{E}$ or $\boldsymbol{H}$. Such condition can be simply stated as that the wave number of the field $\boldsymbol{F}$ satisfies $k \gg \frac{v}{c\lambda}$, where $\lambda$ is the wavelength of the field and $v$ is the moving velocity of the media.

The theory under special relativity can be used to derive the electrodynamics of moving media for a general case, but it may have following technical and even mathematical difficulties. First, special relativity works if the moving speed of the media is a constant and the movement is along a straight line, e.g., an inertial frame, which cannot be easily expanded to cases that the moving velocity is a space and time dependent function, so that the theory may lack of generality. In fact, for varying moving speed cases, there is no simple mathematical expression regarding to Lorentz or Galilean transformation! In general, field theory and special relativity apply only to inertial frame, so that they are unable to easily treat the system that has an acceleration, especially with the invasion of external forces. Secondly, under the assumption of $v \ll c$, Galilean transformation can be an effective and accurate approach for describing the electromagnetic behaviors for applied physics and engineering, as concluded by Rousseaux [13]. These are the considerations being made when we made the first step of the theory for expanded Maxwell's equations [14], which is now more precisely called *Maxwell's equations for mechano-driven slowing-moving media*. In the previous work, we assumed that the moving velocity of the media is a solid translation, which means $\boldsymbol{v}(t)$, the moving velocity and direction depend only on time but not on space coordination. Once there is an external mechanical force input, the media must move and the system is not in an inertia frame, thus, the Maxwell equations may not be Lorentz covariance.

In this paper, we derive the Maxwell's equations for a case in which the media have a time-dependent shapes, volumes and boundaries, and more importantly, the movement of the media is described by a velocity field that is time and space dependent, $\boldsymbol{v}(\boldsymbol{r}, t)$. The only required condition is that the moving speed is much less than the speed of light ($v \ll c$). Such a generalization of the Maxwell's equations will cover the cases in which the media has a non-constant moving velocity, but also the cases of moving charged liquid, fluid or soft charged media system, making it possible to study the electrodynamics of a fluid/liquid matter. Mathematical approaches are proposed for solving the received equations.

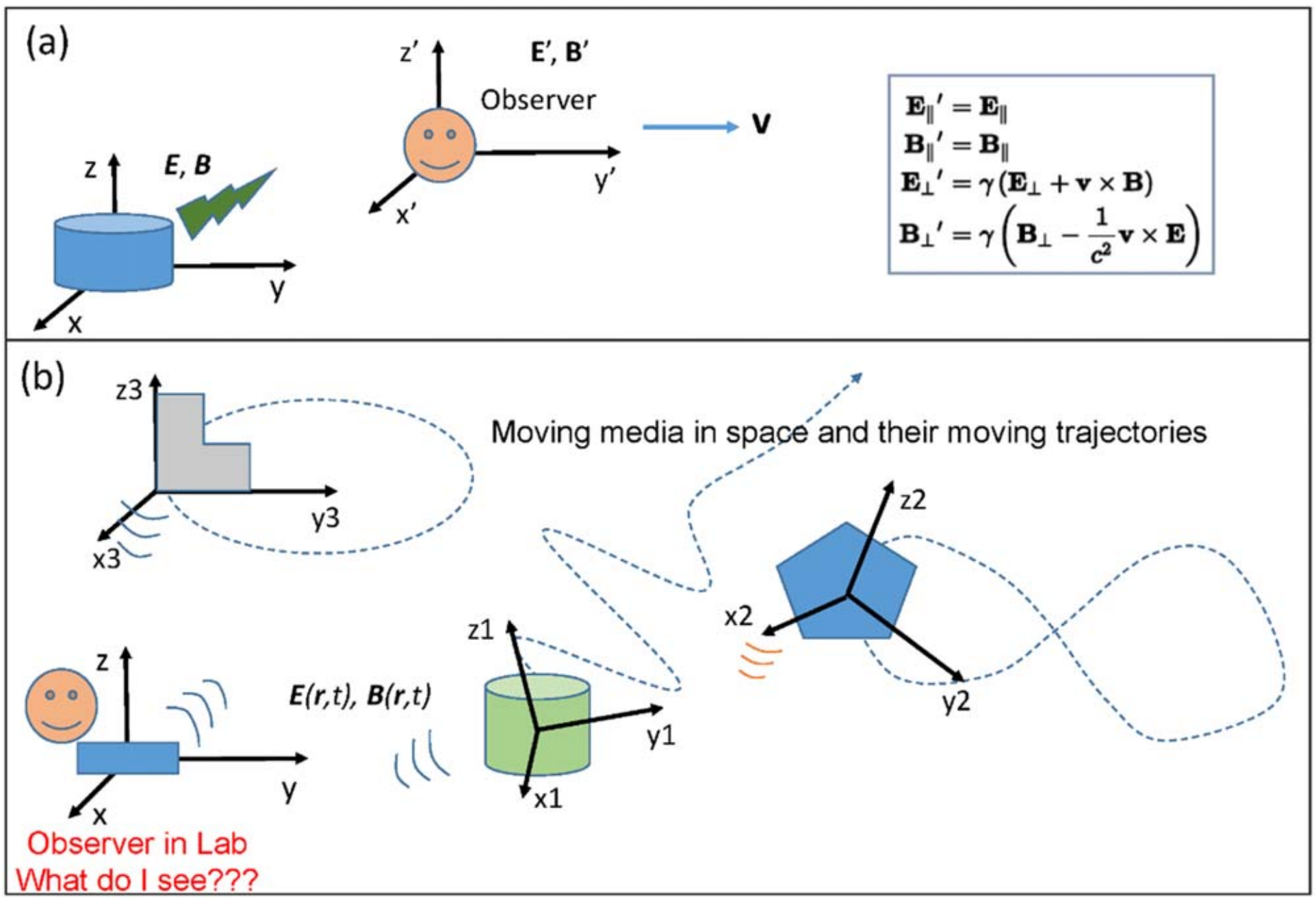

**Figure 1.** (a) Case for special relativity: Schematic diagram representing the general approach in special relativity, in which the fields observed by an observer who is moving at a constant velocity v about the fields ($\mathbf{E}'$), ($\mathbf{B}'$) that was first generated in the Lab frame ($\mathbf{E}$), ($\mathbf{B}$). The Lorentz transformed relationships between the fields in the two reference frame in parallel to $\boldsymbol{v}$ and perpendicular to $\boldsymbol{v}$ are inset. (b) Mechano-driven moving media system: A general case in which the observer is on the ground frame (called Lab frame), with several media moving at complex velocities along various trajectories as represented by the dashed lines. The medium can translate, rotate, expand and even split. The co-moving frames for the media are: (x1, y1, z1), (x2, y2, z2), (x3, y3, z3). In such a case, the Lorentz transformation for special relativity cannot be easily applied, and the only realistic approach is to *express all of the fields in the frame where the observation is done (Lab frame) and all of the fields are expressed in the variables in the same frame.*

## 2. General theory

The traditional approach in field theory of electrodynamics is to derive the Maxwell's equations for stationary medium in Lab frame starting from their integral forms, which are the direct expressions of the four physics laws (see equations (1*a*)–(1*d*)). Then, the equations for a moving medium are derived based on Lorentz transformation from the co-moving frame to the Lab frame as described by special relativity, which assumes that the speed of light in vacuum is a constant in any reference frame and the covariance of the Maxwell's equations in all frames (figure 1(a)). Here, instead of using mathematical transformation between the Lab frame and the co-moving frame, we directly derive the expanded Maxwell's equations from the integral forms of the four physics laws with considering the movement of the medium at the first place. Such derivation may not preserve the Lorentz covariance of the Maxwell's equations, but could be much easier and friendly for practical applications.

The system we are dealing with is much more complex than that in special relativity. In special relativity (figure 1(a)), in which the observer is assumed to move at a constant velocity $\boldsymbol{v}$. In reality, we are dealing with many objects/media that move at a time-dependent velocity and their trajectories can be complex, e.g., non-inertial frame, as schematically shown in figure 1(b). Therefore, we must express the observed fields in the observer frame (Lab frame) for such a case, $\mathbf{E}$ and $\mathbf{B}$ are function of (x, y, z, t), which should be an excellent approximation for the practical system. More importantly, the electromagnetic event in the Lab frame can interact with those in the moving frames so that the entire system needs to be considered systematically and consistently. In such a case, the most convenient and effective approach for describing the electromagnetic fields of the system is in the observer frame (Lab frame).

Our derivations are based on three assumptions: first, the integral forms of the four physics laws must hold (Gauss's law for electricity, Gauss's law for magnetism, Faraday's Electromagnetic induction law (Lenz law), and Ampere-Maxwell law). We start from the integral forms of the physics laws, rather than using the coordination transformation of equations, such as Lorentz transformation or Galilean transformation, both of which are for a medium that moves at a constant velocity along a straight line trajectory. Secondly, we assumed low speed limiting case with $v \ll c$, which is an excellent approximation in engineering and technology. Lastly, we ignore the relativistic effect for the moving media. We anticipate that the results can be more friendly for practical applications. Importantly, for a medium that moves at a varying velocity, there is no simple mathematical expression for Galilean or Lorentz transformation to be applied to the differential equation regarding to coordination transformation.

We first start from the integral forms of the four physics laws that are direct were directly derived from experimentally observed physics phenomena [8, 18]:

$$\oiint_S \boldsymbol{D}' \cdot d\mathbf{s} = \iiint_V \rho_f \, d\boldsymbol{r}.$$

Gauss's law for electricity (1a)

$$\oiint_S \boldsymbol{B} \cdot d\mathbf{s} = 0$$

Gauss's law for magnetism (1b)

$$\oint_C \boldsymbol{E} \cdot d\boldsymbol{L} = -\frac{d}{dt}\iint_C \boldsymbol{B} \cdot d\mathbf{s}$$

Faraday's electromagnetic induction law(Lenz law) (1c)

$$\oint_C \boldsymbol{H} \cdot d\boldsymbol{L} = \iint_C \boldsymbol{J}_f \cdot d\mathbf{s} + \frac{d}{dt}\iint_C \boldsymbol{D}' \cdot d\mathbf{s}$$

Ampere-Maxwell law (1d)

where $\rho_f$ is the density of free charges in space, and $\boldsymbol{J}_f$ is the current density. The surface integrals for $\boldsymbol{B}$ and $\boldsymbol{D}'$ are for a surface that is defined by a closed loop c, and they are the magnetic flux and displacement field flux, respectively. Equation (1a) means that the total electric flux through a closed surface is the total charges enclosed inside the surface. Equation (1b) means that the total magnetic flux through a closed surface is zero. Equation (1c) means that the changing rate of the *total* magnetic flux through an open surface is the induced electric potential around its closed edge loop. Equation (1d) means that the changing rate of the *total* electric flux through an open surface plus the total current flowing across the surface is the integral of the magnetic field around its closed edge loop (electromotive force). The law of the conservation of charges is:

$$\oiint_S \mathbf{J}_f \cdot d\mathbf{s} + \frac{d}{dt}\iiint_V \rho_f \, d\boldsymbol{r} = 0. \quad (2)$$

which means that the total current flux through a closed surface is the changing rate of the total charges enclosed inside the surface. Please note that the sequence of time differentiation and the integral cannot be switched unless for stationary media, because the Maxwell's equations are based on the changing of the *total* magnetic/electric flux, which has to consider both the change in the field and the movement of the media boundaries. For following derivation, we make two important assumptions. First, the integral forms of the four physics laws equations (1a)–(1d) hold for media that have varying shapes and may move in space at an arbitrary low-velocity. Secondly, all of the fields are expressed in the frame where the observation is done, and the coordination and time to be used for describing the fields are defined in this frame as well: $\boldsymbol{E}(\boldsymbol{r}, t)$, $\boldsymbol{B}(\boldsymbol{r}, t)$, $\boldsymbol{H}(\boldsymbol{r}, t)$, $D'(\boldsymbol{r}, t)$, $\rho_f(\boldsymbol{r}, t)$, *and* $J_f(\boldsymbol{r}, t)$. These assumptions are fundamental for our entire theoretical studies.

It is well known that the Lorentz transformation must be used for dealing with Maxwell's equations to warrant their Lorentz covariant, especially when the moving speed of the media is relatively high. However, in practical applications, especially about any mechanical movement on earth, the speed is much less than the speed of light. In such a case, the Galilean transformation could be very accurately used to describe the electromagnetic behavior of these objects although it is not exactly from relativity point of view [19]. Therefore, our discussion here is mainly developing technological applicable theory for moving media in the frame of classical physics.

The most important base of our approach is that *the four physics laws as stated by* equations (1a)–(1e) *hold for moving media in the same reference frame.* This is the key starting point of our following derivation. To derive the differential form of the Maxwell's equations for a moving media whose shape is time-dependent and its velocity field is time-dependent as well, $\boldsymbol{v}(\boldsymbol{r}, t)$, as schematically shown in figure 2. *Our derivation assumes* $v \ll c$, so that the relativistic effect is ignored. For a general field vector $\boldsymbol{C}(\boldsymbol{r},t)$, and its flux through a time-dependent 3D surface defined by $\boldsymbol{s}(t)$, as illustrated in figure A1, is as following:

$$\Phi(t) = \iint_{s(t)} \boldsymbol{C}(\boldsymbol{r}, t) \cdot d\mathbf{s} \quad (3)$$

Please note that the shape, volume of the integral surface is time-dependent. Using the mathematical derivation presented in appendix A, we have:

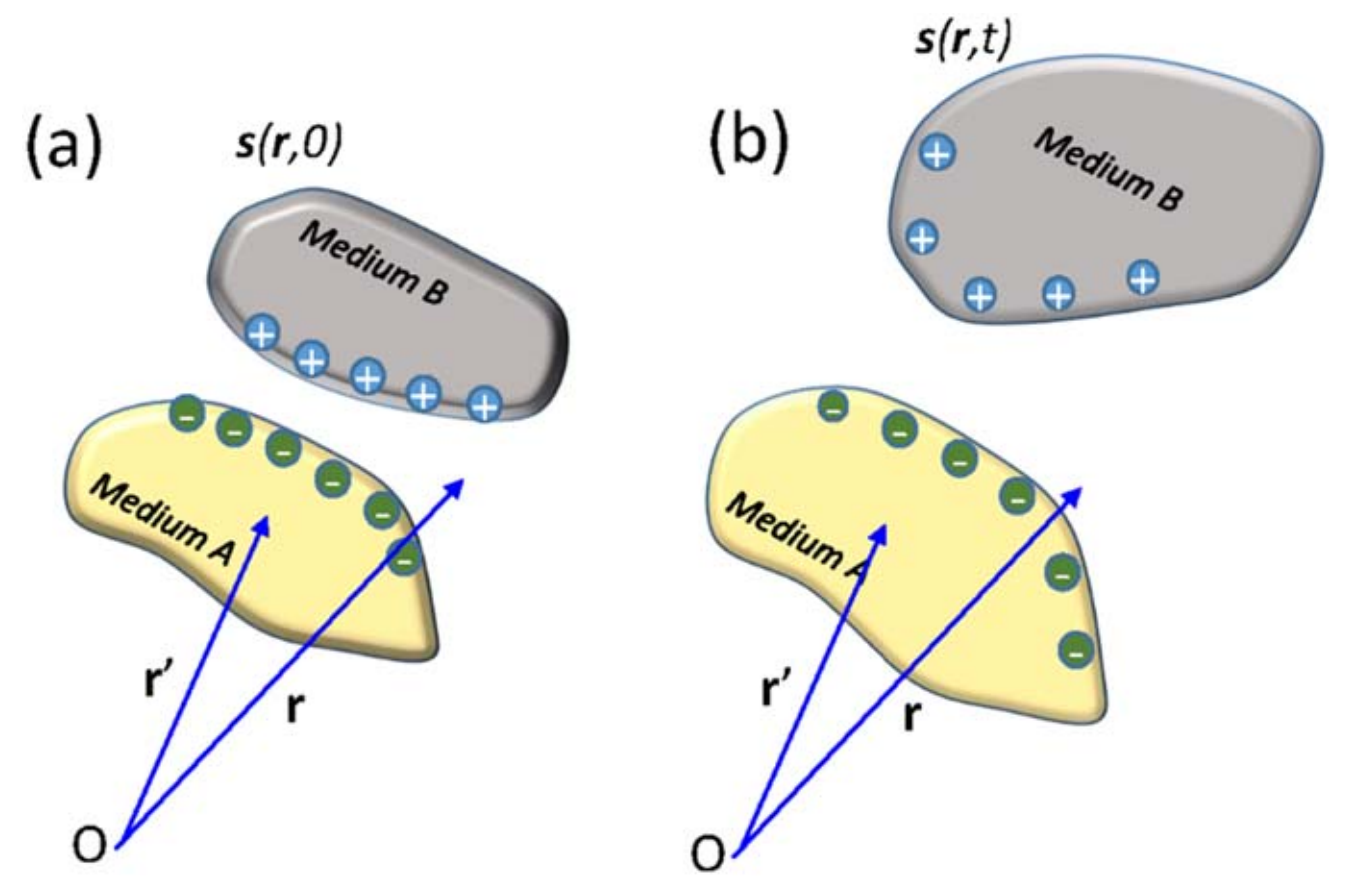

**Figure 2.** Schematic diagram showing a dielectric media system, in which the volume and surfaces of each is time dependent and may move at an arbitrary velocity field. (a) The media distribution at $t = 0$. (b) The media distribution at time $t$. The medium A can be stationary, but medium B is moving. The media can translate, rotate, expand/contraction and even split/combine. The distribution of the electrostatic charges on the surfaces, the boundary and/or volume could also vary owing to the deformation and expansion of the medium surface, such as soft matter and even liquid.

$$\frac{d}{dt}\Phi(t) = \frac{d}{dt}\iint_{s(t)} \boldsymbol{C}(\boldsymbol{r},\, t)\cdot d\boldsymbol{s}$$
$$= \iint_{s(t)} \left\{\frac{\partial}{\partial t}\boldsymbol{C}(\boldsymbol{r},\, t) + [\nabla\cdot\boldsymbol{C}(\boldsymbol{r},\, t)]\boldsymbol{v}(\boldsymbol{r},\, t) - \nabla\times[\boldsymbol{v}(\boldsymbol{r},\, t)\times\boldsymbol{C}(\boldsymbol{r},\, t)]\right\}\cdot d\boldsymbol{s}. \tag{4}$$

Equation (4) is a general mathematical equation for any vector $\boldsymbol{C}$ and medium surface, which will be used in following derivation, in which the local moving velocity $\boldsymbol{v}(\boldsymbol{r},\ t)$ of the medium depends on both time and space, so that it is a general function without restriction. This means that the object can translate, rotate, expand and even split.

In traditional relativity, the fields in Lab frame and in co-moving frame are derived under the Lorentz transformation by warrant the covariance of the Maxwell's equations. In our derivation, we only use one coordination system in which the observer is assumed at stationary, and the media are moving in space; by directly starting from the integral forms of the Maxwell's equations in integral form, we can express what the observer observed without using any coordination transformation, which means that *all of the fields are expressed in the observer's frame and the variables are defined in the same frame as well.* Simply speaking, there is no simple Lorentz transform for a media that moves at an arbitrary velocity field unless one uses general relativity.

Using the mathematical identity as given in equation (4), applying it to equations (1*c*), (1*d*) and use equations (1*a*), (1*b*) and Stokes' s theorem, we have

$$\oiint_s \boldsymbol{D}'\cdot d\boldsymbol{s} = \iiint_V \rho_f\, d\boldsymbol{r}, \tag{5a}$$

$$\oiint_s \boldsymbol{B}\cdot d\boldsymbol{s} = 0, \tag{5b}$$

$$\oint_C \boldsymbol{E}\cdot d\boldsymbol{L} = -\iint_C \frac{\partial}{\partial t}\boldsymbol{B}\cdot d\boldsymbol{s} + \oint_C (\boldsymbol{v}\times\boldsymbol{B})\cdot d\boldsymbol{L}, \tag{5c}$$

$$\oint_C \boldsymbol{H}\cdot d\boldsymbol{L} = \iint_C (J_f + \rho_f\,\boldsymbol{v})\cdot d\boldsymbol{s} + \iint_C \frac{\partial}{\partial t}\boldsymbol{D}'\cdot d\boldsymbol{s} - \oint_C (\,\boldsymbol{v}\times\boldsymbol{D}')\cdot d\boldsymbol{L}. \tag{5d}$$

Equation (5*c*) means that the changing rate of the flux of the magnetic field through an open surface plus the work done by the Lorentz force due to media movement on a unit charge around its edge loop is the induced electric potential drop around its closed edge loop. Equation (5*d*) means that the sum of, the total current flowing through an open surface and the current produced by the free charges due to media movement (right-hand side first term), the changing rate of the electric flux through the open surface (right-hand side second term), and the changing rate of the electric flux produced by media movement (right-hand side third term), is the integral of the magnetic field around its closed edge loop (electromotive force). The terms of $\boldsymbol{v}\times\boldsymbol{B}$ and $\boldsymbol{v}\times\boldsymbol{D}'$ are the sources of produced electromagnetic waves by media movement.

Using the Stokes' s theorem and divergence theorem, we can derive the expanded Maxwell's equations as:

$$\nabla\cdot\boldsymbol{D}'(\boldsymbol{r},\, t) = \rho_f\,(\boldsymbol{r},\, t), \tag{6a}$$

$$\nabla \cdot \boldsymbol{B}(\boldsymbol{r}, t) = 0, \tag{6b}$$

$$\nabla \times [\boldsymbol{E}(\boldsymbol{r}, t) - \boldsymbol{v}(\boldsymbol{r}, t) \times \boldsymbol{B}(\boldsymbol{r}, t)] = -\frac{\partial}{\partial t}\boldsymbol{B}(\boldsymbol{r}, t), \tag{6c}$$

$$\nabla \times [\boldsymbol{H}(\boldsymbol{r}, t) + \boldsymbol{v}(\boldsymbol{r}, t) \times \boldsymbol{D}'(\boldsymbol{r}, t)] = \boldsymbol{J}_f(\boldsymbol{r}, t) + \rho_f(\boldsymbol{r}, t)\boldsymbol{v}(\boldsymbol{r}, t) + \frac{\partial}{\partial t}\boldsymbol{D}'(\boldsymbol{r}, t). \tag{6d}$$

It is important to note that $\boldsymbol{D}'$ is used here to be consistent with our previous notations in [14], rather than representing in co-moving frame. The result is consistent with those derived by Lax and Nelson [9] and Kaufman [20]. Equations (6a)–(6d) may not satisfy the Lorentz covariance, but it is most convenient to treat a general case for slowing moving media especially when the moving velocity is arbitrary and the number of media is more than two (figure 2).

Through the calculus calculation:

$$\frac{d}{dt}\iiint_V \rho_f d\boldsymbol{r} = \iiint_V \left\{\left[\frac{\partial}{\partial t}\rho_f + \boldsymbol{v} \cdot \nabla\rho_f\right]d\boldsymbol{r} + \rho_f \frac{d}{dt}d\boldsymbol{r}\right\}$$
$$= \iiint_V \left[\frac{\partial}{\partial t}\rho_f + \boldsymbol{v} \cdot \nabla\rho_f + \rho_f \nabla \cdot \boldsymbol{v}\right]d\boldsymbol{r} = \iiint_V \left[\frac{\partial}{\partial t}\rho_f + \nabla \cdot (\boldsymbol{v}\rho_f)\right]d\boldsymbol{r}. \tag{7}$$

the law of charge conservation is:

$$\nabla \cdot [\boldsymbol{J}_f(\boldsymbol{r}, t) + \rho_f(\boldsymbol{r}, t)\boldsymbol{v}(\boldsymbol{r}, t)] + \frac{\partial}{\partial t}\rho_f(\boldsymbol{r}, t) = 0. \tag{8}$$

where $\rho_f \boldsymbol{v}$ is the local current produced by the free charges moving at velocity $\boldsymbol{v}$.

Accordingly, from the integral Maxwell's equations equations (5a)–(5d), the boundary conditions can be derived as follows:

$$[\boldsymbol{D}_2' - \boldsymbol{D}_1'] \cdot \boldsymbol{n} = \sigma_f, \tag{9a}$$

$$[\boldsymbol{B}_2 - \boldsymbol{B}_1] \cdot \boldsymbol{n} = 0, \tag{9b}$$

$$\boldsymbol{n} \times [\boldsymbol{E}_2 - \boldsymbol{E}_1 - \boldsymbol{v} \times (\boldsymbol{B}_2 - \boldsymbol{B}_1)] = 0, \tag{9c}$$

$$\boldsymbol{n} \times [\boldsymbol{H}_2 - \boldsymbol{H}_1 + \boldsymbol{v} \times (\boldsymbol{D}_2' - \boldsymbol{D}_1')] = \boldsymbol{K}_s + \sigma_f \boldsymbol{v}_s. \tag{9d}$$

where $\boldsymbol{n}$ is the surface normal direction, $\mathbf{K}_s$ is the surface current density, $\sigma f$ is the surface free charge density, and $\boldsymbol{v}_s$ is the moving velocity of the media in parallel to the boundary.

For moving media, it is possible to have physical contact among the moving media, so that their surfaces must have electrostatic charges due to contact electrification (triboelectric effect) and/or piezoelectric effect. Thus, a variation in medium shape and/or moving medium object results in not only a local time-dependent charge density $\rho s$, but also a local 'virtual' electric current density due to the media movement. To account both terms, the displacement vector is expanded by adding an additional term $\boldsymbol{P}_s$, representing the polarization owing to the pre-existing electrostatic charges on the media; a movement as driven by mechanical action results in a polarization term, so called *mechano-induced polarization*, so that the displacement vector is modified as [21]

$$\boldsymbol{D} = \boldsymbol{D}' + \boldsymbol{P}_s = \varepsilon_0\boldsymbol{E} + \boldsymbol{P} + \boldsymbol{P}_s = \varepsilon_0(1 + \chi)\boldsymbol{E} + \boldsymbol{P}_s. \tag{10}$$

Here, the first term $\varepsilon_0\boldsymbol{E}$ is due to the field created by the free charges, called external electric field; the polarization vector $\boldsymbol{P}$ is the medium polarization caused by the existence of the external electric field $\boldsymbol{E}$; and the added term $\boldsymbol{P}_s$ is mainly due to the existence of the surface electrostatic charges and the time variation in boundary shapes, which were first introduced in order to understand the mechanism of triboelectric nanogenerators and its output power quantification [22–24]. Therefore, equations (6a)–(6d) are expanded as:

$$\nabla \cdot \boldsymbol{D}'(\boldsymbol{r}, t) = \rho_f(\boldsymbol{r}, t) - \nabla \cdot \boldsymbol{P}_s(\boldsymbol{r}, t), \tag{11a}$$

$$\nabla \cdot \boldsymbol{B}(\boldsymbol{r}, t) = 0, \tag{11b}$$

$$\nabla \times (\boldsymbol{E}(\boldsymbol{r}, t) - \boldsymbol{v}(\boldsymbol{r}, t) \times \boldsymbol{B}(\boldsymbol{r}, t)) = -\frac{\partial}{\partial t}\boldsymbol{B}(\boldsymbol{r}, t), \tag{11c}$$

$$\nabla \times [\boldsymbol{H}(\boldsymbol{r}, t) + \boldsymbol{v}(\boldsymbol{r}, t) \times (\boldsymbol{D}'(\boldsymbol{r}, t) + \boldsymbol{P}_s(\boldsymbol{r}, t))]$$
$$= \boldsymbol{J}_f(\boldsymbol{r}, t) + \rho_f(\boldsymbol{r}, t)\boldsymbol{v}(\boldsymbol{r}, t) + \frac{\partial}{\partial t}[\boldsymbol{D}'(\boldsymbol{r}, t) + \boldsymbol{P}_s(\boldsymbol{r}, t)]. \tag{11d}$$

All of the quantities in equations (11a)–(11d) are function of ($\boldsymbol{r}$, $t$) in the Lab frame where the observer is located as in figure 1(b). Equations (11a)–(11d) together with equation (8) are referred as the *general Maxwell's equations for shape-deformable, mechano-driven, slow-moving media at an arbitrary velocity field. The equations describe the coupling among mechanical, electrical and magnetic properties and behaviors of the system*. It needs to point out that we assume that the medium is non-magnetic. An additional term $\nabla \times \boldsymbol{M}$ has to be added in the displacement current if the material has a magnetic moment $\boldsymbol{M}$ [21, 25].

Equation (11) can be viewed as composed of two parts. The first part is the contribution that is proportional to the medium moving velocity $\boldsymbol{v}$, such as $\boldsymbol{v} \times \boldsymbol{B}$ in equation (11c), representing the contribution of the Lorentz force to the local electric field. The term $\boldsymbol{v} \times (\boldsymbol{D}' + \boldsymbol{P}_s)$ in equation (11d) is the local induced electric current due to medium movement to the local electric field. The other part is the $\boldsymbol{v}$ 'independent' terms that are 'as is' in original Maxwell's equations, characterizing the transmission and scattering of electromagnetic waves as well as their interaction with matter. The second part should obey the Lorentz covariance, but the first part may not be, especially for non-inertial frame. Furthermore, the entire equations (11a)–(11d) may not obey the Lorentz covariant even under the condition of $\boldsymbol{v}$ is a constant, e.g., inertial frame, because the mechano-driven term $\boldsymbol{P}_s$ would introduce acceleration/deceleration to the system, representing energy and momentum input from external sources.

To make the calculation converges faster, we can split the velocity field into two components: translation velocity $\boldsymbol{v}_T(t)$, and a local relative movement/rotation velocity $\boldsymbol{v}_r(\boldsymbol{r}, t)$:

$$\boldsymbol{v}(\boldsymbol{r}, t) = \boldsymbol{v}_T(t) + \boldsymbol{v}_r(\boldsymbol{r}, t), \tag{12}$$

where the translation velocity is assumed to be position-independent but does vary with time, and it can be viewed as the moving velocity of the 'center of mass' or 'geometry center' of the media if any. While the relative/rotation velocity is position-dependent, which is about a local relative movement of all the components/parts around the 'center of mass' for example. Equations (11a)–(11d) become:

$$\nabla \cdot \boldsymbol{D}' = \rho_f - \nabla \cdot \boldsymbol{P}_s, \tag{13a}$$

$$\nabla \cdot \boldsymbol{B} = 0, \tag{13b}$$

$$\nabla \times \boldsymbol{E} = -\left(\frac{\partial}{\partial t} + \boldsymbol{v}_T \cdot \nabla\right)\boldsymbol{B} + \nabla \times (\boldsymbol{v}_r \times \boldsymbol{B}), \tag{13c}$$

$$\nabla \times \boldsymbol{H} = \boldsymbol{J}_f + \rho_f \boldsymbol{v}_r + \left(\frac{\partial}{\partial t} + \boldsymbol{v}_T \cdot \nabla\right)(\boldsymbol{D}' + \boldsymbol{P}_s) - \nabla \times [\boldsymbol{v}_r \times (\boldsymbol{D}' + \boldsymbol{P}_s)]. \tag{13d}$$

Equations (13a)–(13d) are the form of the early equations derived based on an assumption that the movement is a solid translation (e.g., $\boldsymbol{v}_r = 0$). The solutions for such a case was presented in [14].

## 3. Conservation of energy as governed by the Maxwell's equations for mechano-driven slow moving media system

Starting from equations (11a)–(11d), we explore the energy conversion process in this mechano-electric-magnetic coupled processes. Using the mathematical identity $\nabla \cdot (\boldsymbol{E} \times \boldsymbol{H}) = \boldsymbol{H} \cdot (\nabla \times \boldsymbol{E}) - \boldsymbol{E} \cdot (\nabla \times \boldsymbol{H})$

And $\boldsymbol{D} = \boldsymbol{D}' + \boldsymbol{P}_s$, we have

$$\begin{aligned}
\iiint_V (\boldsymbol{E} \cdot \boldsymbol{J}_f) d\boldsymbol{r} &= \iiint_V \left(\boldsymbol{E} \cdot \left[\nabla \times \boldsymbol{H} + \nabla \times (\boldsymbol{v} \times \boldsymbol{D}) - \rho_f \boldsymbol{v} - \frac{\partial}{\partial t}\boldsymbol{D}\right]\right) d\boldsymbol{r} \\
&= -\iiint_V \left(\nabla \cdot (\boldsymbol{E} \times \boldsymbol{H}) + \left[\boldsymbol{E} \cdot \frac{\partial \boldsymbol{D}}{\partial t} + \boldsymbol{H} \cdot \frac{\partial \boldsymbol{B}}{\partial t}\right] - \{\boldsymbol{H} \cdot [\nabla \times (\boldsymbol{v} \times \boldsymbol{B})] \right. \\
&\quad \left. + \boldsymbol{E} \cdot [\nabla \times (\boldsymbol{v} \times \boldsymbol{D})]\} + \rho_f \boldsymbol{v} \cdot \boldsymbol{E}\right) d\boldsymbol{r} \\
&= -\oiint_S \boldsymbol{S} \cdot d\boldsymbol{s} - \iiint_V \left(\frac{\partial}{\partial t} u\right) d\boldsymbol{r},
\end{aligned} \tag{14}$$

where $\boldsymbol{S}$ is the Poynting vector

$$\boldsymbol{S} = \boldsymbol{E} \times \boldsymbol{H}, \tag{15a}$$

and $u$ is the energy volume density of electromagnetic field, which is generally given by

$$\frac{\partial}{\partial t} u = \boldsymbol{E} \cdot \frac{\partial \boldsymbol{D}}{\partial t} + \boldsymbol{H} \cdot \frac{\partial \boldsymbol{B}}{\partial t}. \tag{15b}$$

We have

$$\begin{aligned}
&-\iiint_V \left(\frac{\partial}{\partial t} u\right) d\boldsymbol{r} - \oiint_S S \cdot d\boldsymbol{s} \\
&= \iiint_V (\boldsymbol{E} \cdot \boldsymbol{J}') d\boldsymbol{r} + \iiint_V (\rho_f \boldsymbol{v} \cdot \boldsymbol{E}) d\boldsymbol{r} - \iiint_V \{\boldsymbol{H} \cdot [\nabla \times (\boldsymbol{v} \times B)] + \boldsymbol{E} \cdot [\nabla \times (\boldsymbol{v} \times \boldsymbol{D})]\} d\boldsymbol{r},
\end{aligned} \tag{16a}$$

which gives:

$$-\frac{\partial}{\partial t}u - \nabla \cdot \boldsymbol{S} = \boldsymbol{E} \cdot \boldsymbol{J}_f + \rho_f \boldsymbol{v} \cdot \boldsymbol{E} - \{H \cdot [\nabla \times (\boldsymbol{v} \times \boldsymbol{B})] + \boldsymbol{E} \cdot [\nabla \times (\boldsymbol{v} \times (\boldsymbol{D}' + \boldsymbol{P}_s))]\}. \tag{16b}$$

This equation means that the decrease of the internal electromagnetic field energy within a volume plus the rate of electromagnetic wave energy radiated out of the volume surface is the rate of energy done by the field on the external free current and the free charges, plus the media spatial motion induced change in electromagnetic energy density. Therefore, media motion, as triggered by external force, would introduce a non-conservation of electromagnetic wave. Therefore, the Maxwell equations for the mechano-driven media system should not be Lorentz covariant simply due to the energy input by mechanical agitation and media movement.

In equation (16*b*), if we assume that the speed term depends only on time $\boldsymbol{v}(t)$, it can be simplified as

$$-\frac{\partial}{\partial t}u - \nabla \cdot \boldsymbol{S} = \boldsymbol{E} \cdot \boldsymbol{J}_f + \{\boldsymbol{H} \cdot [(\boldsymbol{v} \cdot \nabla)\boldsymbol{B} + \boldsymbol{E} \cdot [(\boldsymbol{v} \cdot \nabla)\boldsymbol{D}']\}. \tag{17a}$$

or equivalently:

$$-\frac{\mathrm{D}}{\mathrm{D}t}u - \nabla \cdot \boldsymbol{S} = \boldsymbol{E} \cdot \boldsymbol{J}_f, \tag{17b}$$

with

$$\frac{\mathrm{D}}{\mathrm{D}t}u = \boldsymbol{E} \cdot \frac{\mathrm{D}\boldsymbol{D}}{\mathrm{D}t} + \boldsymbol{H} \cdot \frac{\mathrm{D}\boldsymbol{B}}{\mathrm{D}t}, \tag{17c}$$

where $\frac{\mathrm{D}}{\mathrm{D}t} = \frac{\partial}{\partial t} + \boldsymbol{v} \cdot \nabla$. At the right-hand side of equation (17*a*), the contribution made by media movement can be considered as 'sources' for producing electromagnetic waves! The media movement could be considered as moving-electromagnetic sources.

## 4. Solution of the expanded Maxwell's equations in time space

We now use the perturbation theory to solve equations (11*a*)–(11*d*) by expanding them in the order of $\lambda$, as a parameter ($\lambda = 1$):

$$\boldsymbol{E} = \boldsymbol{E}_0 + \lambda \boldsymbol{E}_1 + \lambda^2 \boldsymbol{E}_2 + \ldots, \tag{18a}$$

$$D' = \boldsymbol{D}_0' + \lambda \boldsymbol{D}_1' + \lambda^2 \boldsymbol{D}_2' + \ldots, \tag{18b}$$

$$\boldsymbol{H} = \boldsymbol{H}_0 + \lambda \boldsymbol{H}_1 + \lambda^2 \boldsymbol{H}_2 + \ldots, \tag{18c}$$

$$\boldsymbol{B} = \boldsymbol{B}_0 + \lambda \boldsymbol{B}_1 + \lambda^2 \boldsymbol{B}_2 + \ldots \tag{18d}$$

Substituting equations (18*a*), (18*b*) into equations (11*a*)–(11*d*), the corresponding equations for the same order of $\lambda$ are:

For the zeroth order:

$$\nabla \cdot \boldsymbol{D}_0' = \rho_0, \tag{19a}$$

$$\nabla \cdot \boldsymbol{B}_0 = 0, \tag{19b}$$

$$\nabla \times \boldsymbol{E}_0 = -\frac{\partial}{\partial t}\boldsymbol{B}_0, \tag{19c}$$

$$\nabla \times \boldsymbol{H}_0 = \boldsymbol{J}_0 + \frac{\partial}{\partial t}\boldsymbol{D}_0', \tag{19d}$$

where

$$\rho_0 = \rho_f - \nabla \cdot \boldsymbol{P}_s, \tag{19e}$$

$$\boldsymbol{J}_0 = \boldsymbol{J}_f + \rho_f \boldsymbol{v} - \nabla \times (\boldsymbol{v} \times \boldsymbol{P}_s) + \frac{\partial}{\partial t}\boldsymbol{P}_s. \tag{19f}$$

Equations (19*a*)–(19*d*) have the form of classical Maxwell's equations and they can be solved using various methods presented in text books, such as vector potentials, Hertz vectors etc.

For the first order:

$$\nabla \cdot \boldsymbol{D}_1' = 0, \tag{20a}$$

$$\nabla \cdot B_1 = 0, \tag{20b}$$

$$\nabla \times \boldsymbol{E}_1 = \boldsymbol{F}_1 - \frac{\partial}{\partial t}\boldsymbol{B}_1, \tag{20c}$$

$$\nabla \times \boldsymbol{H}_1 = \boldsymbol{J}_1 + \frac{\partial}{\partial t}\boldsymbol{D}_1', \tag{20d}$$

where

$$\boldsymbol{F}_1 = \nabla \times (\boldsymbol{v} \times \boldsymbol{B}_0), \tag{20e}$$

$$\boldsymbol{J}_1 = -\nabla \times (\boldsymbol{v} \times \boldsymbol{D}_0'). \tag{20f}$$

By applying operator $\nabla\times$ to equations (20*c*), (20*d*), we have:

$$\nabla^2\boldsymbol{E}_1 - \mu\varepsilon\frac{\partial^2}{\partial t^2}\boldsymbol{E}_1 = -\nabla \times \boldsymbol{F}_1 + \mu\frac{\partial}{\partial t}\boldsymbol{J}_1, \tag{21a}$$

$$\nabla^2\boldsymbol{H}_1 - \mu\varepsilon\frac{\partial^2}{\partial t^2}\boldsymbol{H}_1 = -\nabla \times \boldsymbol{J}_1 - \varepsilon\frac{\partial}{\partial t}\boldsymbol{F}_1. \tag{21b}$$

Besides the solutions for the homogeneous component, the special solutions $E_{1s}$ and $H_{1s}$ of equations (21*a*), (21*b*) are given as follows:

$$\boldsymbol{E}_{1s}(\boldsymbol{r}, t) = \frac{1}{4\pi}\iiint\frac{1}{|\boldsymbol{r} - \boldsymbol{r}'|}\left[\nabla' \times \boldsymbol{F}_1(\boldsymbol{r}', t') - \mu\frac{\partial}{\partial t'}\boldsymbol{J}_1(\boldsymbol{r}', t')\right]d\boldsymbol{r}', \tag{22a}$$

$$\boldsymbol{H}_{1s}(\boldsymbol{r}, t) = \frac{1}{4\pi}\iiint\frac{1}{|\boldsymbol{r} - \boldsymbol{r}'|}\left[\nabla' \times \boldsymbol{J}_1(\boldsymbol{r}', t') + \mu\frac{\partial}{\partial t'}\boldsymbol{F}_1(\boldsymbol{r}', t')\right]d\boldsymbol{r}'. \tag{22b}$$

where $t'$ is the retardation time $t' = t - \sqrt{\mu\varepsilon}\,|\boldsymbol{r} - \boldsymbol{r}'|$. The total solution has to match the boundary conditions. Please note that the calculation with including the time retardation can be carried out using the method introduced in Jackson's book in Section 6.5 [2].

The second order is:

$$\nabla \cdot \boldsymbol{D}_2' = 0, \tag{23a}$$

$$\nabla \cdot \boldsymbol{B}_2 = 0, \tag{23b}$$

$$\nabla \times \boldsymbol{E}_2 = \boldsymbol{F}_2 - \frac{\partial}{\partial t}\boldsymbol{B}_2, \tag{23c}$$

$$\nabla \times \boldsymbol{H}_2 = \boldsymbol{J}_2 + \frac{\partial}{\partial t}\boldsymbol{D}_2', \tag{23d}$$

where

$$\boldsymbol{F}_2 = \nabla \times (\boldsymbol{v} \times \boldsymbol{B}_1), \tag{23e}$$

$$\boldsymbol{J}_2 = -\nabla \times (\boldsymbol{v} \times \boldsymbol{D}_1'), \tag{23f}$$

By the same token, we have

$$\nabla^2\boldsymbol{E}_2 - \mu\varepsilon\frac{\partial^2}{\partial t^2}\boldsymbol{E}_2 = -\nabla \times \boldsymbol{F}_2 + \mu\frac{\partial}{\partial t}\boldsymbol{J}_2, \tag{24a}$$

$$\nabla^2\boldsymbol{H}_2 - \mu\varepsilon\frac{\partial^2}{\partial t^2}\boldsymbol{H}_2 = -\nabla \times \boldsymbol{J}_2 - \varepsilon\frac{\partial}{\partial t}\boldsymbol{F}_2. \tag{24b}$$

which have the special solution $\boldsymbol{E}_{2s}$ and $\boldsymbol{H}_{2s}$ of:

$$\boldsymbol{E}_{2s}(\boldsymbol{r}, t) = \frac{1}{4\pi}\iiint\frac{1}{|\boldsymbol{r} - \boldsymbol{r}'|}\left[\nabla' \times \boldsymbol{F}_2(\boldsymbol{r}', t') - \mu\frac{\partial}{\partial t'}\boldsymbol{J}_2(\boldsymbol{r}', t')\right]d\boldsymbol{r}', \tag{25a}$$

$$\boldsymbol{H}_{2s}(\boldsymbol{r}, t) = \frac{1}{4\pi}\iiint\frac{1}{|\boldsymbol{r} - \boldsymbol{r}'|}\left[\nabla' \times \boldsymbol{J}_2(\boldsymbol{r}', t') + \mu\frac{\partial}{\partial t'}\boldsymbol{F}_2(\boldsymbol{r}', t')\right]d\boldsymbol{r}'. \tag{25b}$$

Further analytical derivation of equations (25*a*), (25*b*) can follow the Section 6.5 in [2]. The higher orders can be calculated as well. The total solution needs to satisfy the boundary conditions.

## 5. Solution of the expanded Maxwell's equations in frequency space

In general, the dielectric permittivity is frequency dependent, rather than a constant. To include the frequency in the entire theory, we use the Fourier transform and inverse Fourier transform in time and frequency space as defined by:

$$a(\boldsymbol{r}, \omega) = \int_{-\infty}^{\infty} dt\ e^{i\omega t}\ a(\boldsymbol{r}, t), \tag{26a}$$

$$a(\boldsymbol{r},\,t) = \frac{1}{2\pi}\int_{-\infty}^{\infty} d\omega\; e^{-i\omega t}\; a(\boldsymbol{r},\,\omega). \tag{26b}$$

The purpose of introducing frequency space is to simplify the relationship between the displacement field $\boldsymbol{D}'$ and electric field $\boldsymbol{E}$, magnetic field $\boldsymbol{H}$ and magnetic flux density $\boldsymbol{B}$ as follows:

$$\boldsymbol{D}'(\boldsymbol{r},\,\omega) = \varepsilon(\omega)\boldsymbol{E}(\boldsymbol{r},\,\omega), \tag{27a}$$

$$\boldsymbol{B}(\boldsymbol{r},\,\omega) = \mu(\omega)\boldsymbol{H}(\boldsymbol{r},\,\omega). \tag{27b}$$

It is noted that we still use the simplest constitutive relations without considering the corrections made by media movement. Note, we use the same symbols to represent the real space and reciprocal space except the variables. Applying the Fourier transform to equations (19)–(24) and use the perturbation method, we have

The zeroth order:

$$\nabla\cdot\boldsymbol{D}_0'(\boldsymbol{r},\,\omega) = \rho_0(\boldsymbol{r},\,\omega), \tag{28a}$$

$$\nabla\cdot\;\boldsymbol{B}_0(\boldsymbol{r},\,\omega) = 0, \tag{28b}$$

$$\nabla\times\boldsymbol{E}_0(r,\,\omega) = i\omega\boldsymbol{B}_0(\boldsymbol{r},\,\omega), \tag{28c}$$

$$\nabla\times\;\boldsymbol{H}_0(\boldsymbol{r},\,\omega) = \boldsymbol{J}_0(\boldsymbol{r},\,\omega) - i\omega\boldsymbol{D}_0'(\boldsymbol{r},\,\omega). \tag{28d}$$

The first order:

$$\nabla\cdot\boldsymbol{D}_1'(\boldsymbol{r},\,\omega) = 0, \tag{29a}$$

$$\nabla\cdot\boldsymbol{B}_1(\boldsymbol{r},\,\omega) = 0, \tag{29b}$$

$$\nabla\times\;\boldsymbol{E}_1(\boldsymbol{r},\,\omega) = \boldsymbol{F}_1(\boldsymbol{r},\,\omega) + i\omega\boldsymbol{B}_1(\boldsymbol{r},\,\omega), \tag{29c}$$

$$\nabla\times\boldsymbol{H}_1(\boldsymbol{r},\,\omega) = \boldsymbol{J}_1(\boldsymbol{r},\,\omega) - i\omega\boldsymbol{D}_1'(\boldsymbol{r},\,\omega). \tag{29d}$$

The following equations can be derived:

$$\nabla^2\boldsymbol{E}_1(\boldsymbol{r},\,\omega) + \mu\varepsilon\omega^2\boldsymbol{E}_1(\boldsymbol{r},\,\omega) = -\nabla\times\boldsymbol{F}_1(\boldsymbol{r},\,\omega) - i\omega\boldsymbol{J}_1(\boldsymbol{r},\,\omega), \tag{30a}$$

$$\nabla^2\boldsymbol{H}_1(\boldsymbol{r},\,\omega) + \mu\varepsilon\omega^2\boldsymbol{H}_1(\boldsymbol{r},\,\omega) = -\nabla\times\boldsymbol{J}_1(\boldsymbol{r},\,\omega) + i\omega\boldsymbol{F}_1(\boldsymbol{r},\,\omega). \tag{30b}$$

The full solution of the field has two components: homogeneous solution and the special solutions $\boldsymbol{E}_{1s}$ and $\boldsymbol{H}_{1s}$, as given in follows

$$\boldsymbol{E}_{1s}(\boldsymbol{r},\,\omega) = \frac{1}{4}\iiint\frac{\exp\left[i\omega\sqrt{\mu\varepsilon}\,|\mathbf{r}-\mathbf{r}'|\right]}{|\mathbf{r}-\mathbf{r}'|}[\nabla'\times\boldsymbol{F}_1(\boldsymbol{r}',\,\omega) + i\omega\boldsymbol{J}_1(\boldsymbol{r}',\,\omega)]d\boldsymbol{r}', \tag{31a}$$

$$\boldsymbol{H}_{1s}(r,\,\omega) = \frac{1}{4}\iiint\frac{\exp[i\omega\sqrt{\mu\varepsilon}\,|\boldsymbol{r}-\boldsymbol{r}'|]}{|\boldsymbol{r}-\boldsymbol{r}'|}[\nabla'\times\boldsymbol{J}_1(\boldsymbol{r}',\,\omega) - i\omega\boldsymbol{F}_1(\boldsymbol{r}',\,\omega)]d\boldsymbol{r}'. \tag{31b}$$

The second order:

$$\nabla\cdot\boldsymbol{D}_2'(\boldsymbol{r},\,\omega) = 0, \tag{32a}$$

$$\nabla\cdot\boldsymbol{B}_2(\boldsymbol{r},\,\omega) = 0, \tag{32b}$$

$$\nabla\times\;\boldsymbol{E}_2(\boldsymbol{r},\,\omega) = \boldsymbol{F}_2(\boldsymbol{r},\,\omega) + i\omega\boldsymbol{B}_2(\boldsymbol{r},\,\omega), \tag{32c}$$

$$\nabla\times\;\boldsymbol{H}_2(\boldsymbol{r},\,\omega) = \boldsymbol{J}_2(\boldsymbol{r},\,\omega) - i\omega\boldsymbol{D}_2'(\boldsymbol{r},\,\omega). \tag{32d}$$

Similarly:

$$\nabla^2\boldsymbol{E}_2(\boldsymbol{r},\,\omega) + \mu\varepsilon\omega^2\boldsymbol{E}_2(\boldsymbol{r},\,\omega) = -\nabla\times\boldsymbol{F}_2(\boldsymbol{r},\,\omega) - i\omega\boldsymbol{J}_2(\boldsymbol{r},\,\omega), \tag{33a}$$

$$\nabla^2\boldsymbol{H}_2(\boldsymbol{r},\,\omega) + \mu\varepsilon\omega^2\boldsymbol{H}_2(r,\,\omega) = -\nabla\times\boldsymbol{J}_2(\boldsymbol{r},\,\omega) + i\omega\boldsymbol{F}_2(r,\,\omega). \tag{33b}$$

The full solution of the field has two components: homogeneous solution and the special solution solutions $\boldsymbol{E}_{2s}$ and $\boldsymbol{H}_{2s}$, as given in follows

$$\boldsymbol{E}_{2s}(\boldsymbol{r},\,\omega) = \frac{1}{4}\iiint\frac{\exp[i\omega\sqrt{\mu\varepsilon}\,|\mathbf{r}-\mathbf{r}'|]}{|\mathbf{r}-\mathbf{r}'|}[\nabla'\times\boldsymbol{F}_2(\boldsymbol{r}',\,\omega) + i\omega\boldsymbol{J}_2(r',\,\omega)]d\boldsymbol{r}', \tag{34a}$$

$$\boldsymbol{H}_{2s}(\boldsymbol{r},\,\omega) = \frac{1}{4}\iiint\frac{\exp[i\omega\sqrt{\mu\varepsilon}\,|\boldsymbol{r}-\boldsymbol{r}'|]}{|\boldsymbol{r}-\boldsymbol{r}'|}[\nabla'\times\boldsymbol{J}_2(\boldsymbol{r}',\,\omega) - i\omega\boldsymbol{F}_2(\boldsymbol{r}',\,\omega)]d\boldsymbol{r}'. \tag{34b}$$

## 6. Solution in cylindrical coordination system for $v$ = constant case

We now present a solution for a special case in which the media system has a cylindrical symmetry, and the media movement direction is along the symmetry axis, which is defined as the z-axis in our case, $\boldsymbol{v}_T = \boldsymbol{v}_0$ . The second important assumption is that the moving velocity of the media is a constant or slowly varying with time, so that the only the translation component is preserved in equations (13a)–(13d) are simplified as follows:

$$\nabla \cdot \boldsymbol{D}' = \rho_0, \tag{35a}$$

$$\nabla \cdot \boldsymbol{B} = 0, \tag{35b}$$

$$\nabla \times \boldsymbol{E} = -\frac{D}{Dt}\mathbf{B}, \tag{35c}$$

$$\nabla \times \boldsymbol{H} = \boldsymbol{J}_0 + \frac{D}{Dt}\boldsymbol{D}', \tag{35d}$$

where

$$\frac{D}{Dt} = \frac{\partial}{\partial t} + v_0\frac{\partial}{\partial z}. \tag{35e}$$

From equations (35a)–(35d), we can derive a general wave equation that represents either $\boldsymbol{E}$ or $\boldsymbol{H}$ in the form of:

$$\nabla^2\boldsymbol{C}(\mathbf{r}, t) - \mu\varepsilon\frac{D^2}{Dt^2}\boldsymbol{C}(\mathbf{r}, t) = \boldsymbol{G}(\mathbf{r}, t), \tag{36a}$$

$$\text{If } \boldsymbol{C} = \boldsymbol{E},\ \boldsymbol{G} = \frac{\nabla\rho_0}{\varepsilon} + \mu\frac{D}{Dt}\boldsymbol{J}_0; \tag{36b}$$

$$\text{If } \boldsymbol{C} = \boldsymbol{H},\ \boldsymbol{G} = -\nabla \times \boldsymbol{J}_0. \tag{36c}$$

We use the Fourier transform and inverse Fourier transform in time and frequency space as defined by:

$$\boldsymbol{C}(\mathbf{r}, \omega) = \int_{-\infty}^{\infty} e^{i\omega t}C(\mathbf{r}, t)\, dt, \tag{37a}$$

$$\boldsymbol{C}(\mathbf{r}, t) = \frac{1}{(2\pi)}\int_{-\infty}^{\infty} e^{-i\omega t}\boldsymbol{C}(\mathbf{r}, \omega)d\omega. \tag{37b}$$

Equation (36a) becomes:

$$\nabla^2\boldsymbol{C}(\mathbf{r}, \omega) - \mu\varepsilon\frac{D^2}{D\xi^2}\boldsymbol{C}(\mathbf{r}, \omega) = \boldsymbol{G}(\mathbf{r}, \omega), \tag{38a}$$

where

$$\frac{D^2}{D\xi^2} = \left(-i\omega + v_0\frac{\partial}{\partial z}\right)\left(-i\omega + v_0\frac{\partial}{\partial z}\right). \tag{38b}$$

Now let's look at the special solution $\boldsymbol{C}_s$ of equation (38a). We start from its Fourier transform:

$$\boldsymbol{C}_s(k, \omega) = \iiint e^{(-i\boldsymbol{k}\cdot r + i\omega t)}\ \boldsymbol{C}_s(r, \omega)d\boldsymbol{r}, \tag{39a}$$

$$\boldsymbol{C}_s(r, \omega) = \frac{1}{(2\pi)^3}\iiint e^{(i\boldsymbol{k}\cdot\mathrm{r} - i\omega t)}\ \boldsymbol{C}_s(k, \omega)d\boldsymbol{k}. \tag{39b}$$

Substituting equation (39b) into equation (38a), we have

$$-k^2\boldsymbol{C}_s(\boldsymbol{k}, \omega) + \mu\varepsilon(v_0k_z - \omega)^2\boldsymbol{C}_s(k, \omega) = \boldsymbol{G}(k, \omega), \tag{40a}$$

$$\boldsymbol{C}_s(\boldsymbol{k}, \omega) = \frac{\boldsymbol{G}(k, \omega)}{\mu\varepsilon(v_0k_z - \omega)^2 - k^2}. \tag{40b}$$

Substituting equation (40b) into (39b) and (37b), we have

$$\boldsymbol{C}_s(\boldsymbol{r}, t) = \frac{1}{(2\pi)^4}\iiint d\boldsymbol{k}\int_{-\infty}^{\infty} dt\ e^{(i\boldsymbol{k}\cdot\mathrm{r} - i\omega t)}\frac{\boldsymbol{G}(\boldsymbol{k}, \omega)}{\mu\varepsilon\ (v_0k_z - \omega)^2 - k^2}. \tag{41}$$

We now look at the homogeneous solution $\boldsymbol{C}_h$ of equation (38a),

$$\nabla^2\boldsymbol{C}_h(\boldsymbol{r}, \omega) - \mu\varepsilon\left(-i\omega + v_0\frac{\partial}{\partial z}\right)\left(-i\omega + v_0\frac{\partial}{\partial z}\right)\boldsymbol{C}_h(\boldsymbol{r}, \omega) = 0. \tag{42}$$

In the cylindrical coordinator $(\rho, \varphi, \mathrm{z})$, the $j$ component of the $\boldsymbol{C}_h$ satisfies

$$\frac{\partial^2}{\partial\rho^2}C_{hj} + \frac{1}{\rho}\frac{\partial}{\partial\rho}C_{hj} + \frac{1}{\rho^2}\frac{\partial^2}{\partial\varphi^2}C_{hj} + \frac{\partial^2}{\partial z^2}C_{hj} - \mu\varepsilon\left(-i\omega + v_0\frac{\partial}{\partial z}\right)(-i\omega + v_0C_{hj}) = 0, \tag{43a}$$

$$C_{hj}(\boldsymbol{r}, \omega) = R_j(\rho)Q_j(\varphi)Z_j(\mathrm{z}). \tag{43b}$$

Each of which is determined by:

$$\frac{\partial^2}{\partial \varphi^2} Q_j + m^2 Q_j = 0, \tag{44a}$$

$$\frac{\partial^2}{\partial z^2} Z_j - \mu\varepsilon\left(-i\omega + v_0 \frac{\partial}{\partial z}\right)\left(-i\omega + v_0 \frac{\partial}{\partial z}\right) Z_j - K^2 Z_j = 0, \tag{44b}$$

$$\frac{\partial^2}{\partial \rho^2} R_j + \frac{1}{\rho}\frac{\partial}{\partial \rho} R_j - \frac{m^2}{\rho^2} R_j + K^2 R_j = 0. \tag{44c}$$

The general solutions are

$$Q_j(\varphi) = e^{\pm i m \varphi}, \tag{45a}$$

$$Z_j(z) = e^{\pm \Lambda z}, \tag{45b}$$

where $\Lambda$ is determined by:

$$\Lambda^2 - \mu\varepsilon\,(-i\omega \mp v_0 \Lambda)^2 - K^2 = 0. \tag{45c}$$

The solutions of $R_j(\rho)$ are given by the Bessel functions. The final solutions are the sum of the special solution with a linear superposition of the homogeneous solutions, and the superposition coefficients are determined by matching the total solutions with the boundary conditions as given by equations (9a)–(9d).

## 7. Maxwell's equations including constitutive relations for $v$ = constant case

In general derivation, the constitutive relations are usually assumed to be $\boldsymbol{D}' = \varepsilon \boldsymbol{E}$ and $\boldsymbol{B} = \mu \boldsymbol{H}$ for linear media, which are simple for calculations but they are valid only for stationary media. For moving media, we have to use the following constitutive relations [13]:

$$\boldsymbol{D}' = \varepsilon \boldsymbol{E} - \varepsilon \boldsymbol{v}_0 \times \boldsymbol{B}, \tag{46a}$$

$$\boldsymbol{H} = \boldsymbol{B}/\boldsymbol{\mu} + \varepsilon \boldsymbol{v}_0 \times \boldsymbol{E}. \tag{46b}$$

Substituting equation (42a) into equations (31a)–(31d), and only keep the first order term of $\boldsymbol{v}_0$, we have

$$\varepsilon \nabla \cdot \boldsymbol{E} = \rho_0 + \varepsilon\, \nabla \cdot (\boldsymbol{v}_0 \times \boldsymbol{B}), \tag{47a}$$

$$\nabla \cdot \boldsymbol{B} = 0, \tag{47b}$$

$$\nabla \times (E - \boldsymbol{v}_0 \times \boldsymbol{B}) = -\frac{\partial}{\partial t}\boldsymbol{B}, \tag{47c}$$

$$\begin{aligned}
\nabla \times (\boldsymbol{B}/\mu + \varepsilon \boldsymbol{v}_0 \times \boldsymbol{E}) &= \boldsymbol{J}_0 + \left(\frac{\partial}{\partial t} + \boldsymbol{v}_0 \cdot \nabla\right)(\varepsilon \boldsymbol{E} - \varepsilon\, \boldsymbol{v}_0 \times \boldsymbol{B}) \\
&\cong \boldsymbol{J}_0 + \varepsilon\left(\frac{\partial}{\partial t} + \boldsymbol{v}_0 \cdot \nabla\right)\boldsymbol{E} - \varepsilon \boldsymbol{v}_0 \times \frac{\partial}{\partial t}\boldsymbol{B} \\
&\cong \boldsymbol{J}_0 + \varepsilon \frac{\partial}{\partial t}\boldsymbol{E} + \varepsilon[(\boldsymbol{v}_0 \cdot \nabla)\boldsymbol{E} + \boldsymbol{v}_0 \times \nabla \times \boldsymbol{E}] \\
&= \boldsymbol{J}_0 + \varepsilon \frac{\partial}{\partial t}\boldsymbol{E} + \varepsilon \nabla(\boldsymbol{v}_0 \cdot \boldsymbol{E}),
\end{aligned}$$

$$\nabla \times (\boldsymbol{B}/\mu + \varepsilon \boldsymbol{v}_0 \times \boldsymbol{E}) \cong \boldsymbol{J}_0 + \varepsilon \nabla(\boldsymbol{v}_0 \cdot \boldsymbol{E}) + \varepsilon \frac{\partial}{\partial t}\boldsymbol{E},$$

Or

$$\nabla \times \boldsymbol{B} \cong \mu \boldsymbol{J}_0 + \mu\varepsilon\left(\frac{\partial}{\partial t} + \boldsymbol{v}_0 \cdot \nabla\right)\boldsymbol{E}, \tag{47d}$$

which are the new set of Maxwell's equations derived for a medium that is slowly moving at a constant velocity. By applying $\nabla\ \cdot$ on equation (47c), we can prove that the law of conservation of charges is automatically obeyed:

$$\nabla \cdot (\boldsymbol{J}_0 + \boldsymbol{v}_0 \rho_0) + \frac{\partial}{\partial t}\rho_0 = 0. \tag{47e}$$

By the same token, equations (11a)–(11d) could be modified as following for a constant moving speed case:

$$\varepsilon \nabla \cdot \boldsymbol{E} = \rho_f - \nabla \cdot \boldsymbol{P}_s + \varepsilon \nabla \cdot (\boldsymbol{v}_0 \times \boldsymbol{B}), \tag{48a}$$

$$\nabla \cdot \boldsymbol{B} = 0, \tag{48b}$$

$$\nabla \times (E - v_0 \times B) = -\frac{\partial}{\partial t}B, \tag{48c}$$

$$\nabla \times (B/\mu + \varepsilon v_0 \times E) = J_f + \left(\frac{\partial}{\partial t} + v_0 \cdot \nabla\right)(\varepsilon E - \varepsilon v_0 \times B + P_s)$$

$$\cong J_f + \varepsilon\left(\frac{\partial}{\partial t} + v_0 \cdot \nabla\right)E - \varepsilon\, v_0 \times \frac{\partial}{\partial t}B + \left(\frac{\partial}{\partial t} + v_0 \cdot \nabla\right)P_s$$

$$\cong J_f + \varepsilon\frac{\partial}{\partial t}E + \varepsilon[(v_0 \cdot \nabla)E + v_0 \times \nabla \times E] + \left(\frac{\partial}{\partial t} + v_0 \cdot \nabla\right)P_s$$

$$= J_f + \varepsilon\frac{\partial}{\partial t}E + \varepsilon\nabla(v_0 \cdot E) + \left(\frac{\partial}{\partial t} + v_0 \cdot \nabla\right)P_s,$$

$$\nabla \times [B/\mu + v_0 \times (\varepsilon E + P_s)] = J_f + \varepsilon\nabla(v_0 \cdot E) + \frac{\partial}{\partial t}(\varepsilon E + P_s),$$

$$\nabla \times B = \mu J_0 + \mu\frac{\partial}{\partial t}P_s + \mu\varepsilon\left(\frac{\partial}{\partial t} + v_0 \cdot \nabla\right)E. \tag{48d}$$

Equations (48*a*)–(48*d*) are the *Maxwell's equations for mechno-driven slowing moving media system*. The term $v_0 \cdot E$ in equation (48*d*) may be related to the power required to move the medium, and its gradient is likely to be the current produced by the mechano-driving. This term also appear in the equations give in [17]. Equations (48*a*)–(48*d*) can be solved using the mathematical methods systematically introduced in [14] in both time and frequency spaces.

## 8. Relativistic corrections for $v = $ constant case

The equations we have derived are equivalent to the Hertz equations except that the media movement velocity is a general function, which are the non-relativistic (low velocity) limit of Lorentz transformation of Maxwell's equations. Previously, if one considers the correction to be made by relativistic effect, and if $v$ is a constant $v_0$, an $\alpha$ factor should be added in front of $v_0 \times B$ and $v_0 \times D'$ terms in equations (6*c*)–(6*d*) [8], thus:

$$\nabla \cdot D' = \rho_f, \tag{49a}$$

$$\nabla \cdot B = 0, \tag{49b}$$

$$\nabla \times (E - \alpha v_0 \times B) = -\frac{\partial}{\partial t}B, \tag{49c}$$

$$\nabla \times (H + \alpha v_0 \times D') = J_f + \alpha\rho_f v_0 + \frac{\partial}{\partial t}D' \tag{49d}$$

The $\alpha$ factor represents the corrections made by relativistic effect for the case that $v$ is a constant [15]:

$$\alpha = 1 - \frac{\mu_0\varepsilon_0}{\mu\varepsilon}. \tag{49e}$$

However, as for our case that $v$ is a field that is a function of position and time, we may adopt the $\alpha$ factor correction for special relativity case, but its mathematical proof remains to be validated if this procedure is physically reasonable. If it is valid, equations (11*a*)–(11*d*) could be modified as:

$$\nabla \cdot D' = \rho_f - \nabla \cdot P_s, \tag{50a}$$

$$\nabla \cdot B = 0, \tag{50b}$$

$$\nabla \times (E - \alpha v \times B) = -\frac{\partial}{\partial t}B, \tag{50c}$$

$$\nabla \times [H + \alpha v \times (D' + P_s)] = J_f + \alpha\rho_f v + \frac{\partial}{\partial t}(D' + P_s). \tag{50d}$$

The law of conservation of charges is:

$$\nabla \cdot [J_f + \alpha\rho_f v] + \frac{\partial}{\partial t}\rho_f = 0. \tag{50e}$$

Accordingly, the Maxwell's equation for a mechano-driven system equations (48*a*)–(48*d*) are modified as:

$$\varepsilon\nabla \cdot E = \rho_f - \nabla \cdot P_s + \varepsilon\alpha\nabla \cdot (v_0 \times B), \tag{51a}$$

$$\nabla \cdot B = 0, \tag{51b}$$

$$\nabla \times (\boldsymbol{E} - \alpha \boldsymbol{v}_0 \times \boldsymbol{B}) = -\frac{\partial}{\partial t}\boldsymbol{B}, \tag{51c}$$

$$\nabla \times [\boldsymbol{B}/\mu + \alpha \boldsymbol{v}_0 \times (\varepsilon \boldsymbol{E} + P_s)] = \boldsymbol{J}_f + \varepsilon\alpha\nabla(\boldsymbol{v}_0 \cdot E) + \frac{\partial}{\partial t}(\varepsilon \boldsymbol{E} + \boldsymbol{P}_s). \tag{51d}$$

Here we do have a question. In the conventional Lorentz transformation, the speed of light in the term $v/c_0$ is always taken as the speed of light in vacuum ($c_0 = (\mu_0\varepsilon_0)^{-1/2}$), based on which the $\alpha$ factor was derived [15]. In the case for a dielectric medium, should $c = (\mu\varepsilon)^{-1/2}$ be taken as the speed of light in the media that is to be used in the Lorentz transformation? We are not sure about the answer. If this is the case, we have to reconsider the expression of the $\alpha$ factor added in here [8, 15]. We leave it as an open question.

## 9. Comparisons of approaches toward electromagnetic theories for moving media

In the relativistic theory, space and time are unified. Under this theoretical scheme, for media moving at a constant speed along a straight line, special relativity can be easily applied to this case without assuming low-moving speed limit. However, for a media system that the media move along complex and in non-inertial frame, which means that the speed is a function of time at least, the theory for general relativity may be required for such a case. Although such an approach is rigorous, but its application for engineering purposes is too complex to be easily carried out. Also, for the speed we care about for objects on earth, the corrections from general relativity should be very minimal small and can be ignored.

In the Galilean absolute space and time scheme, there are two approaches. One is the Galilean electromagnetism [13], in which the field in space at a time t is taken as a quasi-static case, e.g. the 'frozen' field assumption. Therefore, the media distribution and related fields are treated 'frame by frame' (as in films for a movie) under quasi-static approximation. The theory of moving media can be treated frame by frame with the use of constitutive relationships under slow-media moving cases. More approximations can be made for magnetic-dominated or electric-dominated systems. Again, such theory can be easily applied if the moving speed is a constant along a straight line, but we have to check if it can be applied for complex media moving trajectory cases, in which the dynamics may not be fully considered and the relativistic effect is ignored. Again, we are not sure if the constitutive relationship one has proposed holds in non-inertial frame, in which the speed depends on time and space.

The second approach in Galilean absolute space and time scheme is the theory we presented here using the Maxwell equations for a mechano-driven slow moving media system, which can be applied to any media that move along complex trajectories in non-inertial frame as long as the moving speed is low and the relativistic effect is ignored. Such equations should not be Lorentz covariant simply due to the energy input from mechanical triggering. This approach is more effective for applied physics, which has been widely used in engineering electrodynamics [26–28]. The conditions under which the Galilean transformation stands as following [29]:

1. The relative speed between two inertial frames of reference is much smaller than the speed of light in vacuum: $v_0 \ll c$; and
2. Galilean phenomenon takes place in an arena, the spatial extension of which is much smaller than the distance traveled by light during the duration of the phenomenon: $x \ll ct$.

We believe that the above two conditions are valid at least for the electromagnetic phenomena that we are care about on earth.

## 10. Conjunction of the Maxwell equations for a mechano-driven system and the standard Maxwell equations in space

Now we consider a case for a media system that have several moving media along complex trajectories, and our observation is done on the Laboratory frame (see figure 3 ). For the space inside each media, the governing equations are the Maxwell equations for a mechano-driven system (e.g. equations (11*a*)–(11*d*), which apparently do not obey the Lorentz covariance. In this case, one does not have to worry about the exceeding of the speed of light in vacuum $c_0$, because the speed of the light inside media $c_m$ is always slower than $c_0$, and $v \ll c_m$. Once the electromagnetic wave is generated, its traveling outside of the media is governed by the Maxwell equations for vacuum, which of course are Lorentz covariance, which means the speed of light remains constant regardless the media is moving or not. The solutions of the both sets of equations meet at the media boundaries as governed by boundary conditions equation (9).

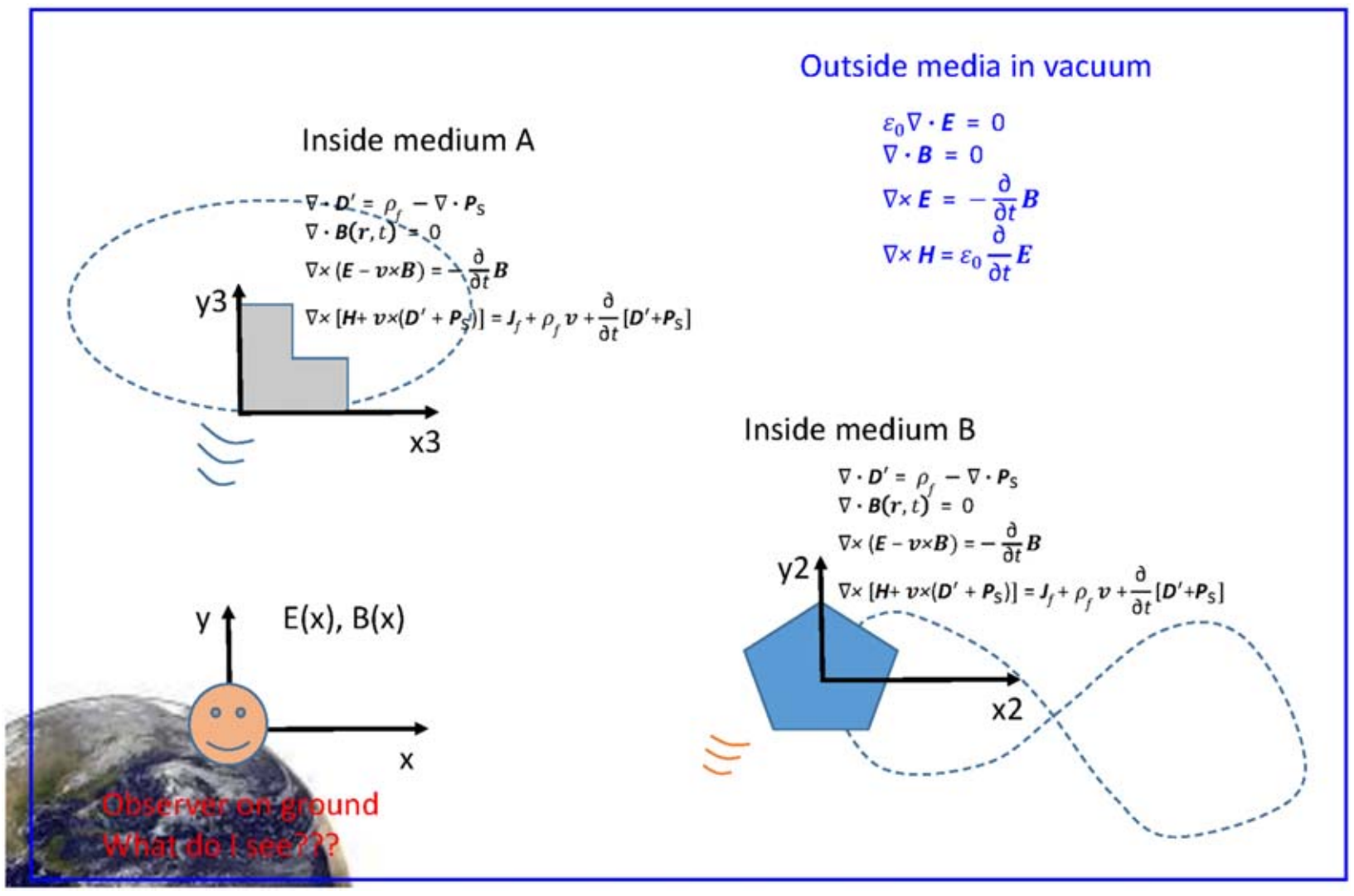

**Figure 3.** Conjunction of the Maxwell equations for a mechano-driven system and the standard Maxwell equations as the media are moving along the dashed trajectories, and the observation is done on earth (in Lab frame).

## 11. Discussions and conclusions

It is generally believed that the description of electromagnetic wave for a moving media should start with the special relativity, which assumes that the moving velocity $v$ is a constant and the movement trajectory is a straight line. In a case that the speed is time and position dependent, the theory would be too complicated to be stated analytically in any simple way. Although special relativity is the formal theory under Lorentz covariance that is beautiful and perfect only if the moving speed is a constant and the moving trajectory is a straight line. As for practical applications, approximations have to be made to make the mathematical derivation reasonably easy but does not loss the accuracy in any measurable way. Therefore, although the expanded Maxwell's equations presented here were derived under the low-moving speed limiting case, it is clear, concise, reliable, self-consistent, sufficient accurate and ready to be applied in many fields. In our daily life, 9 times of speed of sound is $\sim$3 km s$^{-1}$, which is 100,000th of the speed of light. We believe that the results derived under Galilean transformation should be accurate enough for describing the physics experiments made on earth using macroscopic objects, unless one considers the movement of elementary particles, such as electrons and neutrons.

In conclusion, starting from the integral form of the Maxwell's equations, we have derived the Maxwell's equations in differential form for a mechano-driven media system that has a time-dependent shape and volume and may move at an arbitrary velocity field by assuming $v \ll$ c. The equation includes the coupling among mechanical force—electricity and magnetism, and it may not satisfy Lorentz covariance simply because the total energy of electricity and magnetism is not conservative under the input of external mechanical energy. The equations should be applicable to not only moving charged solid and soft media, but also applicable to charged fluid/liquid media, so that it is possible to describe the electrodynamics for fluid/liquid media if the fluid mechanics is included for predicting the media shape and movement velocity field. General strategies for solving the expanded Maxwell's equations are presented using the perturbation theory.

The current theory can be expanded to other systems. Magnetohydrodynamics is a field that is about the magnetic behavior of electrically conductive fluids, such as plasmas, liquid metals, salt water and electrolytes [30]. Besides the electrodynamics, the theory for fluid also include the mass and charge transport equations and Cauchy momentum equation for flowing fluid [31]. Furthermore, recent studies show that electron transfer and ion transfer do occur at liquid-solid interface [32, 33] and even liquid-liquid interface [34]. This means that there are charges distributing at and in liquid media owing to contact-electrification effect. How to accurately describe the electrodynamic behavior of fluid/liquid and liquid-like matter requires innovative theoretical approach that involves not only fluid dynamics, but also *fluid electrodynamics*. Therefore, a set of equations can be set up that not only include the mass-transport and energy transport equations in fluid dynamics, but also include the Maxwell's equations for mechano-driven moving media presented here. The current theory should provide an approach for such a system at least from the electrodynamics point of view.

Studying the electrodynamics of moving media has broad applications. Besides the fundamental theory for triboelectric nanogenerator [21] that may operate up to MHz-GHz frequencies, studying the mechano-driven electromagnetic radiation could be very interesting for wireless signal transmission. Wireless communication for moving objects, such as train, cars and aircrafts, can be revisited for improving the stability of signals. Lastly, high resolution images are usually formed using phase information of the scattered wave, such as electron microscopy, and optical holography. We wonder if the phase information of electromagnetic wave could be used for high resolution radar imaging. This should very possible because electromagnetic wave is light. We believe that the theory presented here should set the theoretical foundation for these applications.

At the end, there are several interesting questions that remain to be discussed and investigated:

1. For a stationary medium, the differential form of the Maxwell's equations preserves the Lorentz covariance, but we are not sure if the integral form of the Maxwell's equations (equation (5)) would preserve the Lorentz covariance. If not, what happened when the integral form was converted into the differential form? How does the covariance be changed if we introduce the moving medium boundaries in equation (1)? One may argue that electromagnetic wave (photon) has no mass, so that it has to be described using Lorentz transformation. But the moving media have mass and they are realistic objects, it can be described using Galilean space and time.

2. The standard differential form of the Maxwell equations is Lorentz covariance, which is the beauty of the theory and it could coincidently happened over 160 years ago when Maxwell composed the Maxwell's equations [1]. As of today, shall we understand that the Maxwell's equations were coined absolutely to be Lorentz covariance, otherwise one cannot simply call it Maxwell's equations? is this the case? We may not fully agree. We need to point out the Lorentz covariance is only valid for inertia frame. For a system that has mechano-driven media movement, the system is not an inertia system, thus, the derived equations should not satisfy Lorentz covariance. Therefore, the equations we have derived in the current work (such as equation (11) in this paper and equation (15) in [14]) still be called expanded Maxwell's equations.

3. For a medium that moves along a straight line at a constant velocity, e.g. inertial frame, the electromagnetic fields in the Lab frame and the co-moving frame can be correlated by the Lorentz transformation (see figure 1(a)). If the moving velocity of the medium, both in magnitude and/or direction, varies with time, do we have to treat the electromagnetic behavior of such a system using the general relativity even for a slow moving medium? If yes, this will be too complex to be done in practice for engineering purposes. In such a case, we believe that our approach is probably more feasible. Our approach resonances with the conclusion of Rousseaux [13, 35]: 'The Galilean limits provide an efficient way to analyze a large amount of phenomena studied by both engineers and physicists.'

4. The Maxwell's equations are covariance in all inertial reference frames, how about non-inertial frames, such as the case we discussed here, how about the system that has a mechano-driven process? We should not generalize the Lorentz covariance preserved under special relativity for inertial frame to general cases, simply because of the introduction of mechano-driven term and the medium movement under force with acceleration. In practice, almost all of the cases we are dealing with have medium movement with acceleration/deceleration, e.g., non-inertial frame. Therefore, can we develop a realistic approach that may not obey the Lorentz covariance exactly but do solve the problems we care about in applied physics and engineering! This is the objective of the current work, which may be different from the point of view of theoretical physicists who are more keen on the preservation of the Lorentz covariance of the Maxwell's equations, so that the beauty and symmetry of physics are preserved, possibly resulting in the future unification of four forces in universe. However, in applied physics and engineering, we believe that usefulness is more important than beautifulness! This is the guiding idea for us to develop the theory presented in this work and in [14].

5. As Rousseaux [13] pointed out in 2013: 'One century after the seminal work of Minkowski, the electrodynamics of moving continuous media is still a subject of investigations for research and should be included in physics lectures as early as possible'. Our teaching on electrodynamics is still based on text books probably developed over 50 years ago. The only lectures on moving media are about special relativity in inertial frame. There are few lectures related to Galilean electromagnetism that is more practical and realistic for general cases, so that generations of students may not have the background knowledge about the forefront development in this classical field. We suggest that there should be a change about the contents of text books in electrodynamics so that the teaching materials can be up to the most recent advances.

6. Since the Faraday's law of electromagnetic induction holds exactly only in inertia frame, the standard differential form of the Maxwell's equations is exact in inertia frame of references in which the media are at rest, based on which the standard field theory has been established. This could mean that the field theory holds in inertia frame of references. However, for non-inertial frame in which the media may move with an acceleration, the Maxwell's equations may need to be expanded. But such a topic is rarely discussed in literature and remains as a topic to be investigated. Since there is no exact inertia frame of references existing on earth and in universe, such an assumption is only theoretical ideal case, but not perfectly satisfied in practice. One cannot always using general relativity for engineering electromagnetism, which is too complex to have sensitive solutions. Therefore, the Maxwell's equations for a mechano-driven slow-moving media system we have developed are a step forward toward the electrodynamics in non-inertia frame of references. We anticipate its future applications in many fields.

All data that support the findings of this study are included within the article.

## Acknowledgments

Thanks to Professors Jin-Min Yang, Fei Wang, Qun Wang, Jiajia Shao and Xu Guo for stimulation discussions.

## Data availability statement

The data that support the findings of this study are available upon reasonable request from the authors.

## Appendix A

We consider a general mathematical calculation of a general vector $\boldsymbol{C}(\boldsymbol{r})$ through a surface defined by $\boldsymbol{s}(t)$ in 3D space as illustrated in figure A1 [36].

$$\Phi(t) = \iint_{s(t)} \boldsymbol{C}(\boldsymbol{r}, t) \cdot d\boldsymbol{s}. \tag{A1}$$

To calculate the time differentiation of the flux $(t)$, we first consider a small increment $\Delta t$ on time $t$, and separate the differentiation over the field and the surface over which the integral is to be carried out:

$$\begin{aligned} \Phi(t + \Delta t) - \Phi(t) &= \iint_{s(t+\Delta t)} \boldsymbol{C}(\boldsymbol{r}, t + \Delta t) \cdot d\boldsymbol{s} - \iint_{s(t)} \boldsymbol{C}(\boldsymbol{r}, t) \cdot d\boldsymbol{s} \\ &= \iint_{s(t+\Delta t)} [\boldsymbol{C}(\boldsymbol{r}, t + \Delta t) - \boldsymbol{C}(\boldsymbol{r}, t)] \cdot d\boldsymbol{s} + \delta, \end{aligned} \tag{A2a}$$

where

$$\delta = \iint_{s(t+\Delta t)} [\boldsymbol{C}(\boldsymbol{r}, t)] \cdot d\boldsymbol{s} - \iint_{s(t)} \boldsymbol{C}(\boldsymbol{r}, t) \cdot d\boldsymbol{s}. \tag{A2b}$$

Using the model shown in figure A1, the surface integral for a full volume $V(t) + \Delta V$ minuses the integral of that for the initial volume $V(t)$, where $\Delta V$ is the volume of a thin shell above $V$ due to the media movement as indicated, we have

$$\left\{\iint_{s(t+\Delta t)} [\boldsymbol{C}(\boldsymbol{r}, t)] \cdot d\boldsymbol{s} + \iint_{L(t)} \boldsymbol{C}(\boldsymbol{r}, t) \cdot [d\boldsymbol{L} \times \boldsymbol{v}(\boldsymbol{r}, t)\Delta t]\right\} - \iint_{s(t)} \boldsymbol{C}(\boldsymbol{r}, t) \cdot d\boldsymbol{s}, \tag{A3}$$

where the second term is an integral over the ring-belt surface at the bottom surface as a result of the media movement. The loop for defining the ring-belt is $\boldsymbol{L}$, and the side length is $\boldsymbol{v}(\boldsymbol{r}, t)\Delta t$, so that the surface integral unit is $(d\boldsymbol{s} = d\boldsymbol{L} \times \boldsymbol{v}(\boldsymbol{r}, t)\Delta t)$. Using the definition in equation (A2b) and the Stokes's theorem, considering that the common sharing of the bottom disk surface in figure A1, by converting the surface integral for a closed surface into a volume integral, we have

$$\begin{aligned} \delta + \iint_{L(t)} \boldsymbol{C}(\boldsymbol{r}, t) \cdot [d\boldsymbol{L} \times \boldsymbol{v}(\boldsymbol{r}, t)\Delta t] &= \iiint_{V+\Delta V} \nabla \cdot \boldsymbol{C}(\boldsymbol{r}, t) d\boldsymbol{r} - \iiint_{V} \nabla \cdot \boldsymbol{C}(\boldsymbol{r}, t) d\boldsymbol{r} \\ &= \iiint_{\Delta V} \nabla \cdot \boldsymbol{C}(\boldsymbol{r}, t) d\boldsymbol{r} = \iiint_{\Delta V} \nabla \cdot \boldsymbol{C}(\boldsymbol{r}, t)\, (\boldsymbol{v}(\boldsymbol{r}, t)\Delta t \cdot d\boldsymbol{s}), \end{aligned} \tag{A4}$$

where $(\boldsymbol{v}(\boldsymbol{r}, t)\Delta t) \cdot d\boldsymbol{s}$ is the volume unit for the shell. Solving $\delta$ from equation (A4),

$$\begin{aligned} \delta &= \iiint_{V} \nabla \cdot \boldsymbol{C}(\boldsymbol{r}, t)\, (\boldsymbol{v}(\boldsymbol{r}, t)\Delta t \cdot d\boldsymbol{s}) - \iint_{L(t)} \boldsymbol{C}(\boldsymbol{r}, t) \cdot [d\boldsymbol{L} \times \boldsymbol{v}(\boldsymbol{r}, t)\Delta t] \\ &= \iiint_{V} \nabla \cdot \boldsymbol{C}(\boldsymbol{r}, t)\, \boldsymbol{v}(\boldsymbol{r}, t) \cdot d\boldsymbol{s}\Delta t - \iint_{L(t)} \boldsymbol{v}(\boldsymbol{r}, t) \times \boldsymbol{C}(\boldsymbol{r}, t) \cdot d\boldsymbol{L}\Delta t. \end{aligned} \tag{A5}$$

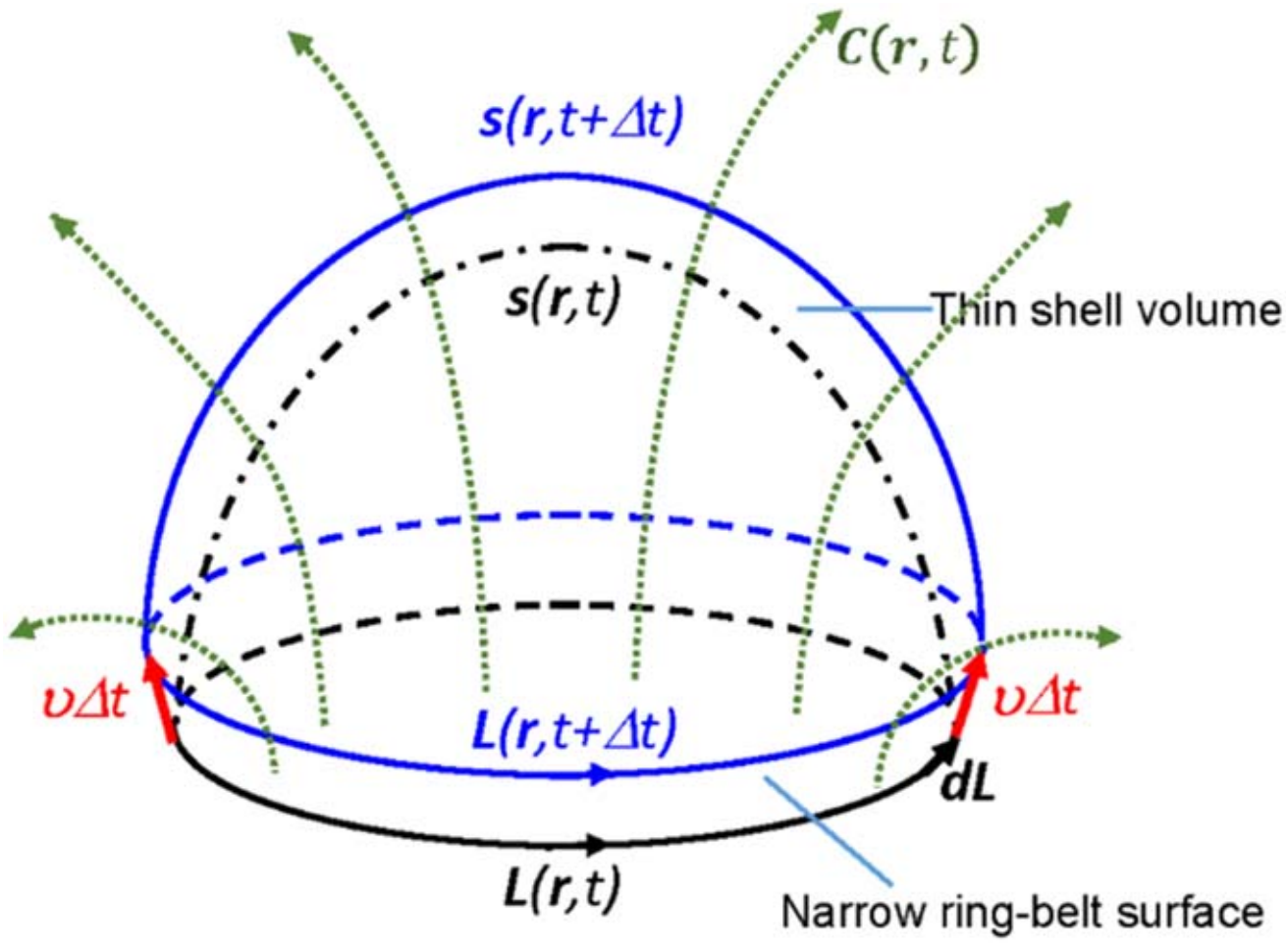

**Figure A1.** Schematic diagram showing the movement of a volume and its surface from time t to time t + Δt, where the thin shell volume unit and the belt-like surface area unit will be used for the mathematical derivation.

Substituting equation (A5) into (A2*a*), we have

$$\frac{d}{dt}\Phi(t) = \frac{d}{dt}\iint_{s(t)} \boldsymbol{C}(\boldsymbol{r}, t)\cdot d\boldsymbol{s} = \lim_{\Delta t\to 0}\frac{\Phi(t+\Delta t) - \Phi(t)}{\Delta t}$$
$$= \iint_{s(t)}\left\{\frac{\partial}{\partial t}\boldsymbol{C}(\boldsymbol{r}, t) + [\nabla\cdot\boldsymbol{C}(\boldsymbol{r}, t)]\boldsymbol{v}(\boldsymbol{r}, t) - \nabla\times[\boldsymbol{v}(\boldsymbol{r}, t)\times\boldsymbol{C}(\boldsymbol{r}, t)]\right\}\cdot d\boldsymbol{s}. \quad \text{(A6)}$$

Equation (A6) is a general mathematical equation for any vector $\boldsymbol{C}$ and medium surface, which will be used in following derivation.

## ORCID iDs

Zhong Lin Wang https://orcid.org/0000-0002-5530-0380